\documentclass[aps,prd,superscriptaddress,twoside,twocolumn,nofootinbib,showpacs,floatfix]{revtex4-1}
\usepackage{amsmath,amssymb}
\usepackage{graphicx}
\usepackage{subfigure}
\usepackage{braket}
\usepackage[colorlinks]{hyperref}
\usepackage[usenames,dvipsnames]{color}
\pdfpagewidth=\paperwidth
\pdfpageheight=\paperheight
\allowdisplaybreaks

\begin{document}

\title{\boldmath Analysis of the data on spin density matrix elements for $\gamma p \to K^{*+}\Lambda$}

\author{Neng-Chang Wei}
\affiliation{Department of Physics, Zhengzhou University, Zhengzhou, Henan 450001, China}
\affiliation{School of Nuclear Science and Technology, University of Chinese Academy of Sciences, Beijing 100049, China}

\author{Ai-Chao Wang}
\affiliation{School of Nuclear Science and Technology, University of Chinese Academy of Sciences, Beijing 100049, China}

\author{Fei Huang}
\email{huangfei@ucas.ac.cn}
\affiliation{School of Nuclear Science and Technology, University of Chinese Academy of Sciences, Beijing 100049, China}

\author{De-Min Li}
\email{lidm@zzu.edu.cn}
\affiliation{Department of Physics, Zhengzhou University, Zhengzhou, Henan 450001, China}

\date{\today}

\begin{abstract}
In our previous work [Phys. Rev. C {\bf 96}, 035206 (2017)], the high-precision differential cross-section data for $\gamma p \to K^{*+}\Lambda$ reported by the CLAS Collaboration has been analyzed within an effective Lagrangian approach. It was found that apart from the $t$-channel $K$, $K^*$, and $\kappa$ exchanges, the $u$-channel $\Lambda$, $\Sigma$, and $\Sigma^*$ exchanges, the $s$-channel $N$ exchange, and the interaction current, one needs to introduce at least two nucleon resonances in the $s$ channel in constructing the reaction amplitudes to describe the cross-section data. One of the needed resonances is $N(2060)5/2^-$, and the other one could be one of the $N(2000)5/2^+$, $N(2040)3/2^+$, $N(2100)1/2^+$, $N(2120)3/2^-$, and $N(2190)7/2^-$ resonances. In this paper, we further include in our analysis the data on spin density matrix elements for $K^*$ meson reported recently by the CLAS Collaboration, with the purpose being to impose further constraints on extracting the resonance contents and to gain a better understanding of the reaction mechanism. It turns out that with the new data on spin density matrix elements taken into account, only the set with the $N(2060)5/2^-$ and $N(2000)5/2^+$ resonances among those five possible solutions extracted from the analysis of the differential cross-section data can satisfactorily describe the data on both the differential cross sections and the spin density matrix elements. Further analysis shows that this reaction is dominated by the $t$-channel $K$ exchange and $s$-channel $N(2060)5/2^-$ and $N(2000)5/2^+$ exchanges.
\end{abstract}



\maketitle

\section{Introduction}   \label{Sec:intro}

Photoproduction of mesons other than pion off a nucleon is believed to be suitable to study the so-called missing resonances,  i.e., the nucleon resonances that are predicted by the quark model calculations \cite{Isgur:1977ef, Koniuk:1979vy} or lattice QCD simulations \cite{Edwards:2011, Edwards:2013, Engel:2013, Lang:2013, Lang:2017, Kiratidis:2017, Andersen:2018} but escaped from experimental detection in $\pi N$ scattering or pion photoproduction reactions, as these resonances may couple weakly to $\pi N$ but strongly to other baryon-meson states. Not only that, the $K^*\Lambda$ photoproduction has the following advantages in studying the excited nucleon states ($N^*$'s): (i) it has a better chance than pion production reactions to reveal resonances with sizable hidden $s\bar{s}$ content as both the $K^*$ and $\Lambda$ have non-zero strangeness; (ii) it is more suited than pion production reactions to investigate the high mass $N^*$'s as $K^*\Lambda$ has a much higher threshold; and (iii) it acts as an isospin filter excluding the contributions of $\Delta^*$'s as $K^*\Lambda$ has isospin $I=1/2$.

The first high-precision differential cross-section data for the reaction $\gamma p \to K^{*+}\Lambda$ were reported by the CLAS Collaboration at the Thomas Jefferson National Accelerator Facility (JLab) in 2013 \cite{Tang:2013}. Theoretically, these data have been analyzed in Refs.~\cite{Wang:2017tpe,Wang:2019,Kim:2014hha} within effective Lagrangian approaches and in Ref.~\cite{Yu:2016} within a Regge model. We mention that before the differential cross-section data were published in 2013 \cite{Tang:2013}, several theoretical and experimental works had already been devoted to the studies of the $\gamma p \to K^{*+}\Lambda$ reaction \cite{Zhao:2001jw,Guo:2006kt,Oh:2006hm,Oh:2006in,Ozaki:2009wp,Hicks:2010pg,Kim:2011rm}, as has been briefly reviewed in Ref.~\cite{Wang:2017tpe}.

It is well known that the cross-section data alone is far from being sufficient to uniquely determine the resonance contents in meson production reactions. In our previous work [Phys. Rev. C {\bf 96}, 035206 (2017)], a satisfactory description of the differential cross-section data for $\gamma p \to K^{*+}\Lambda$ has been achieved within an effective Lagrangian approach. There, it was found that apart from the $t$-channel $K$, $K^*$, and $\kappa$ exchanges, the $u$-channel $\Lambda$, $\Sigma$, and $\Sigma^*$ exchanges, the $s$-channel $N$ exchange, and the interaction current, in order to get a satisfactory description of the differential cross-section data, at least two nucleon resonances should be introduced in constructing the $s$-channel reaction amplitudes. One of the needed resonances is the $N(2060)5/2^-$, while the other one cannot be uniquely determined. It could be any one of the $N(2000)5/2^+$, $N(2040)3/2^+$, $N(2100)1/2^+$, $N(2120)3/2^-$, and $N(2190)7/2^-$ resonances. One thus got five fits with roughly similar fitting qualities, all visually in good agreement with the cross-section data. It is expected that the data on spin observables would provide more constraints and help to distinguish these five different fits.

Recently the data on spin density matrix elements for $\gamma p \to K^{*+}\Lambda$ have been reported by the CLAS Collaboration in Ref.~\cite{Anisovich:2017rpe}, where a partial wave analysis from BnGa group is also presented. It is natural to expect that these new data will impose additional constraints besides those imposed by the differential cross-section data alone in constructing the reaction amplitudes for $\gamma p \to K^{*+}\Lambda$. Nevertheless, as far as we know, these new data have so far never been analyzed in any theoretical calculations, especially in the framework of effective Lagrangian approaches or Regge model.

In the present work, we for the first time include in our analysis the new data on spin density matrix elements for $K^*$ meson within an effective Lagrangian approach based on our previous work \cite{Wang:2017tpe}. The purpose is to further pin down the resonance contents and the associated resonance parameters and to get a better understanding of the reaction mechanism for this reaction.

This paper is organized as follows. In the next section, we briefly introduce the framework of our model. The numerical results of cross sections and spin density matrix elements are shown and discussed in Sec.~\ref{Sec:results}. Finally, a brief summary and conclusions are given in Sec.~\ref{sec:summary}.

\section{Formalism}  \label{Sec:formalism}

\begin{figure}[tbp]
\subfigure[]{
\includegraphics[width=0.45\columnwidth]{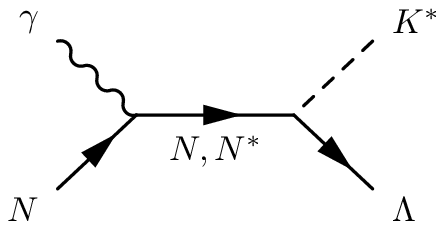}}  {\hglue 0.4cm}
\subfigure[]{
\includegraphics[width=0.45\columnwidth]{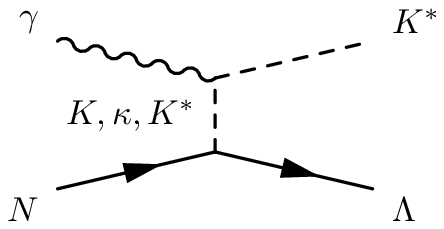}} \\[6pt]
\subfigure[]{
\includegraphics[width=0.45\columnwidth]{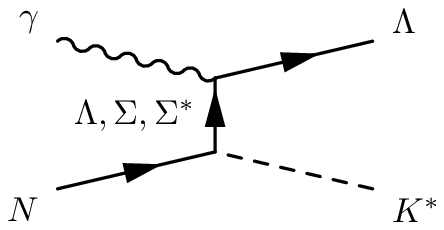}} {\hglue 0.4cm}
\subfigure[]{
\includegraphics[width=0.45\columnwidth]{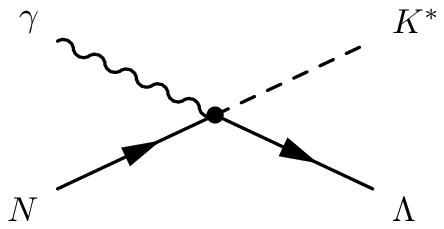}}
\caption{Generic structure of the $K^*$ photoproduction amplitude for $\gamma N\to K^{*}\Lambda$. Time proceeds from left to right. (a) $s$channel, (b) $t$ channel, (c) $u$ channel, and (d) Interaction current.}
\label{FIG:feymans}
\end{figure}

According to a field theoretical approach of Refs.~\cite{Haberzettl:1997,Haberzettl:2006,Huang:2012,Huang:2013}, the full photoproduction amplitude for $\gamma N \to K^* \Lambda$ can be expressed as \cite{Wang:2017tpe}
\begin{equation}
M^{\nu\mu} = M^{\nu\mu}_s + M^{\nu\mu}_t + M^{\nu\mu}_u + M^{\nu\mu}_{\rm int},  \label{eq:amplitude}
\end{equation}
where $\nu$ and $\mu$ indicate the indices of vector meson $K^*$ and photon $\gamma$, respectively. $M^{\nu\mu}_s$ stands for the $s$-channel amplitude, which includes the contributions from the $N$ and $N^\ast$'s exchanges. $M^{\nu\mu}_t$ represents the $t$-channel amplitude, which includes the contributions from the $\kappa$, $K$, and $K^*$ exchanges. $M^{\nu\mu}_u$ stands for the $u$-channel amplitude, which includes the contributions from the $\Lambda$, $\Sigma$, and $\Sigma^\ast$ exchanges. These three terms of the amplitudes can be obtained by direct evaluations of the corresponding Feynman diagrams, which are schematically depicted in Fig.~\ref{FIG:feymans}. The last term $M^{\nu\mu}_{\rm int}$ in Eq.~(\ref{eq:amplitude}) represents the interaction current, which arises from the photon attaching inside the $\Lambda N K^*$ interaction vertex. A strict calculation of $M^{\nu\mu}_{\rm int}$ is impractical as it is in principle highly nonlinear and includes very complicated diagrams. We follow Refs.~\cite{Haberzettl:1997,Haberzettl:2006,Huang:2012,Huang:2013} to choose a particular prescription for $M^{\nu\mu}_{\rm int}$, which obeys the crossing symmetry and ensures that the full photoproduction amplitude for $\gamma N \to K^* \Lambda$ satisfies the generalized Ward-Takahashi identity and thus is fully gauge invariant.

The explicit expressions of the Lagrangians, propagators, and form factors needed for the calculation of $M^{\nu\mu}_s$, $M^{\nu\mu}_t$, and $M^{\nu\mu}_u$ and the adopted prescription for $M^{\nu\mu}_{\rm int}$ can be found in our previous work \cite{Wang:2017tpe}, and we do not repeat them here.

In the center-of-mass (c.m.) frame, the $\gamma p \to K^{*+} \Lambda$ invariant amplitude $M^{\nu\mu}$ introduced in Eq.~(\ref{eq:amplitude}) can be expressed in helicity basis as \cite{Jacob:1964}
\begin{equation}
T_{\lambda_V\lambda_f\lambda_\gamma\lambda_i} \!\! \left(W,\theta\right) \equiv \Braket{{\boldsymbol q},\lambda_V; {\boldsymbol p}_f,\lambda_f | M | {\boldsymbol k},\lambda_\gamma; {\boldsymbol p}_i,\lambda_i},     \label{eq:HelMtrx}
\end{equation}
where $\lambda_i$, $\lambda_\gamma$, $\lambda_f$, and $\lambda_V$ denote the helicities of the incoming nucleon, incoming photon, outgoing $\Lambda$, and outgoing $K^\ast$, respectively. Correspondingly, the arguments ${\boldsymbol p}_i$, ${\boldsymbol k}$, ${\boldsymbol p}_f$, and ${\boldsymbol q}$ represent the momentum of the incoming nucleon, incoming photon, outgoing $\Lambda$, and outgoing $K^*$, respectively. $W$ and $\theta$ indicate the total energy of the system and the scattering angle in c.m. frame, respectively. The differential cross section is then given by
\begin{equation}
\frac{d\sigma}{d\Omega} = \frac{1}{64\pi^2 W^2} \frac{|{\boldsymbol q}|}{|{\boldsymbol k}|} \frac{1}{4} \sum_{\lambda_V\lambda_f \lambda_\gamma \lambda_i} \left | T_{\lambda_V\lambda_f\lambda_\gamma\lambda_i} \!\!\left(W,\theta\right)\right|^2,   \label{eq:3-1}
\end{equation}
and one of the spin density matrix elements, $\rho^0$, which is relevant to the present work, is given by \cite{Wei:2019,Schilling:1969um}
\begin{equation}
\rho^0_{\lambda_V \lambda'_V}  = \frac{\sum_{\lambda_f \lambda_\gamma \lambda_i} T_{\lambda_V\lambda_f \lambda_\gamma\lambda_i}\!\!\left(W,\theta\right) {T_{\lambda'_V\lambda_f\lambda_\gamma\lambda_i}\!\!\left(W,\theta\right)}^* }{\sum_{\lambda_V\lambda_f \lambda_\gamma \lambda_i} \left | T_{\lambda_V\lambda_f\lambda_\gamma\lambda_i}\!\!\left(W,\theta\right) \right|^2}.  \label{eq:SDMEs}
\end{equation}
In the following parts of the paper, the superscript 0 of the spin density matrix elements will be omitted everywhere for the sake of conciseness.

\section{Results and discussion}   \label{Sec:results}

\begin{figure*}[tbp]
\includegraphics[width=0.65\textwidth]{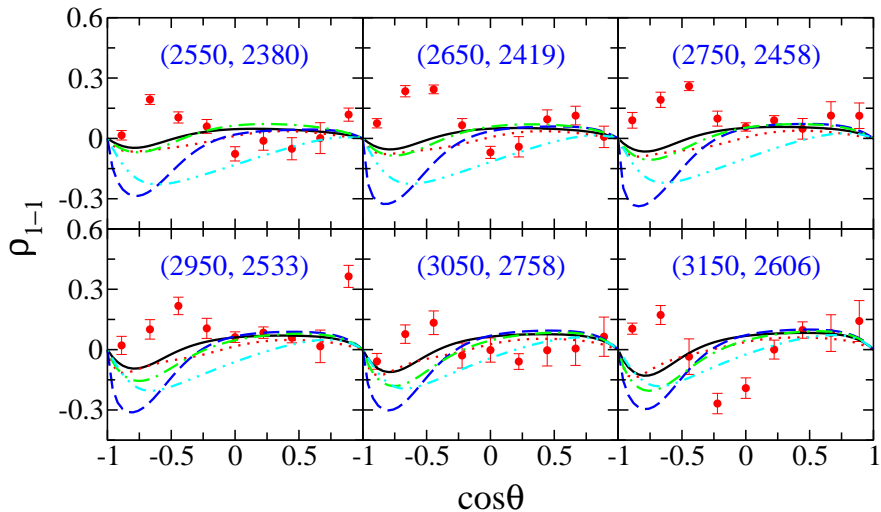}
\caption{Predictions of $\rho_{1-1}$ for $\gamma p \to K^{*+}\Lambda$ from the five fits reported in Ref.~\cite{Wang:2017tpe}. The numbers in parentheses, respectively, denote the photon laboratory incident energy (left number) and the total center-of-mass energy of the system (right number), in MeV. The red dotted, green dash-dotted, black solid, blue dashed, and cyan dash-double-dotted lines correspond to the predictions from fits 1--5 of Ref.~\cite{Wang:2017tpe}, respectively. The scattered symbols are the recent data from the CLAS Collaboration \cite{Anisovich:2017rpe}. }
\label{fig:ac}
\end{figure*}

\begin{figure}[tbp]
\vglue 0.25cm
\includegraphics[width=\columnwidth]{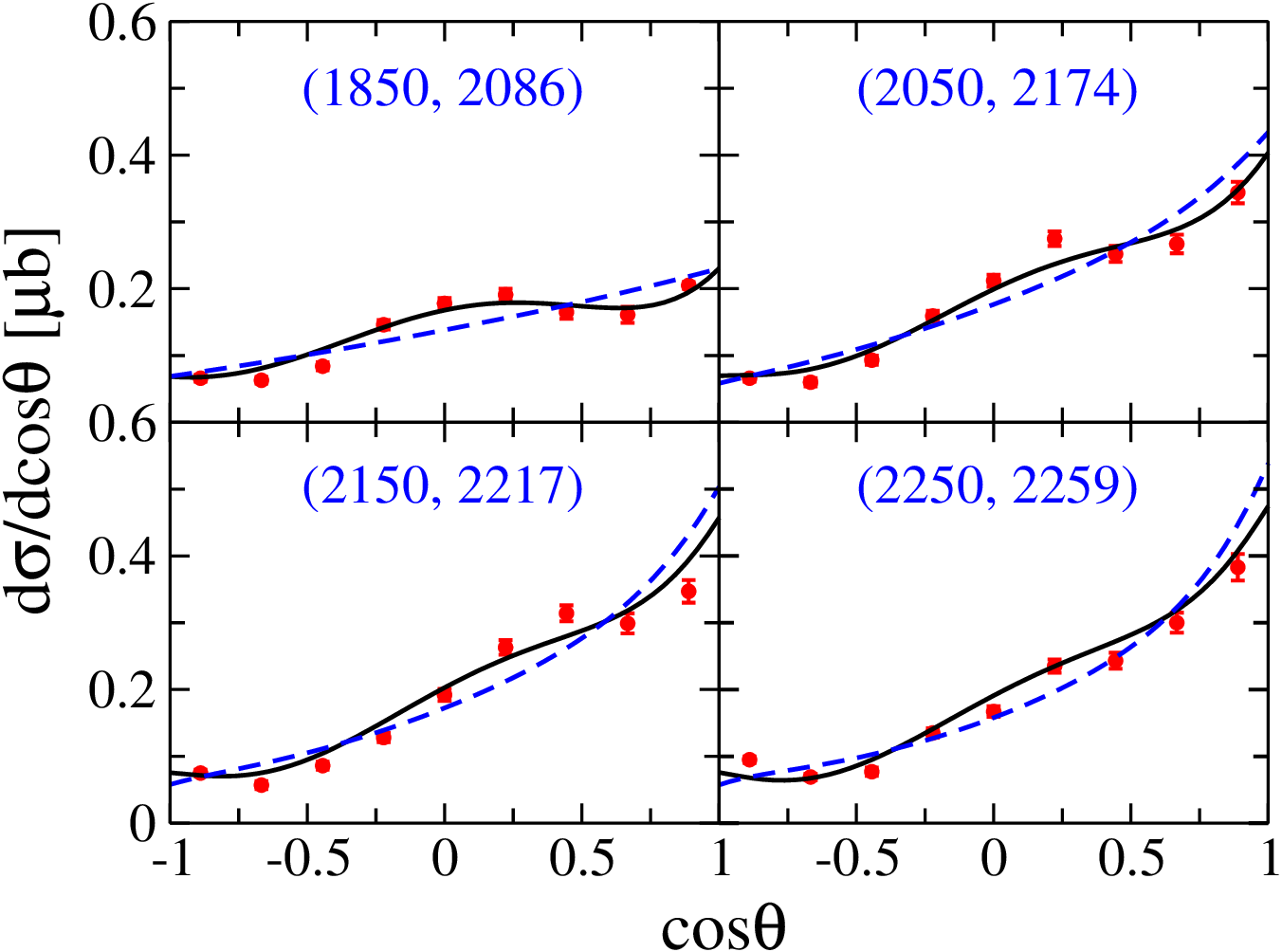}
\caption{Differential cross sections for $\gamma p \to K^{*+}\Lambda$ as a function of $\cos\theta$ in the center-of-mass frame. The black solid lines represent the results with the inclusion of the $N(2000)5/2^+$ and $N(2060)5/2^-$ resonances. The blue dashed lines denote the results with the inclusion of the $N(2040)3/2^+$ and $N(2120)3/2^-$ resonances. The scattered symbols are data from the CLAS Collaboration \cite{Tang:2013}. The numbers in parentheses, respectively, denote the photon laboratory incident energy (left number) and the total center-of-mass energy of the system (right number), in MeV.}
\label{fig:compare}
\end{figure}

\begin{table*}[htb]
\caption{\label{table:chi2} $\chi^2_i/N_i$ evaluated for a given type of observable specified by the index $i=d\sigma$ (differential cross section), $\rho_{00}$, $\rho_{1-1}$, and ${\rm Re}\,\rho_{10}$, with $N_i$ being the corresponding number of data points considered. In the second column, the numbers in the brackets denote the corresponding values for the data in the energy range $W\leq 2217$ MeV. The last column corresponds to the global $\chi^2/N$, where $N$ is the total number of data points including all the types of observables considered.}
\renewcommand{\arraystretch}{1.2}
\begin{tabular*}{\textwidth}{@{\extracolsep\fill}lccccc}
\hline\hline
      & $\chi^2_{d\sigma}/N_{d\sigma}$ & $\chi^2_{\rho_{00}}/N_{\rho_{00}}$  & $\chi^2_{\rho_{1-1}}/N_{\rho_{1-1}}$  & $\chi^2_{\rho_{10}}/N_{\rho_{10}}$ & $\chi^2/N$  \\
 &$N_{d\sigma}$=191 (45) & $N_{\rho_{00}}$=180 & $N_{\rho_{1-1}}$=180 & $N_{\rho_{10}}$=180 & $N$=731 \\
\hline
$N(2040)3/2^+$, $N(2120)3/2^-$   & 2.4 (2.0)  & 5.6   & 6.2  & 5.4 & 4.9\\
$N(2000)5/2^+$, $N(2060)5/2^-$   & 2.8 (1.2)  & 6.3   & 5.3  & 6.9 & 5.3 \\
$N(2040)3/2^+$, $N(2060)5/2^-$   & 3.2 (1.9)  & 9.1   & 6.3  & 4.6 & 5.8 \\
$N(2120)3/2^-$, $N(2060)5/2^-$   & 3.4 (2.6)  & 9.0  & 5.0  & 6.2 & 5.9\\
$N(2190)7/2^-$, $N(2060)5/2^-$   & 2.3 (1.3)   & 5.4  & 7.5  & 9.4 & 6.1  \\
\hline\hline
\end{tabular*}
\end{table*}

\begin{figure*}[tbp]
\includegraphics[width=0.65\textwidth]{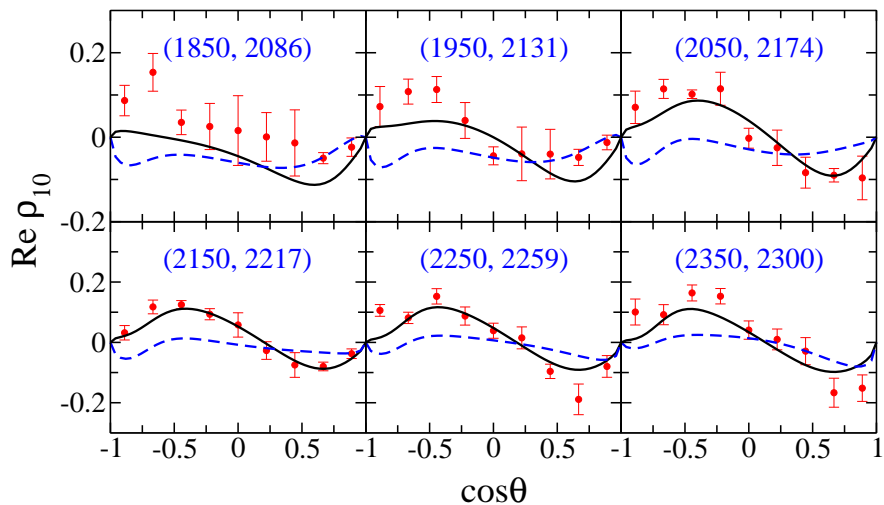}
\caption{Real parts of $\rho_{10}$ for $\gamma p \to K^{*+}\Lambda$ as a function of $\cos\theta$ in the center-of-mass frame. The black solid lines represent the results with the inclusion of the $N(2000)5/2^+$ and $N(2060)5/2^-$ resonances. The blue dashed lines denote the results with the inclusion of the $N(2190)7/2^-$ and $N(2060)5/2^-$ resonances. The scattered symbols are the recent data from the CLAS Collaboration \cite{Anisovich:2017rpe}. The numbers in parentheses, respectively, denote the photon laboratory incident energy (left number) and the total center-of-mass energy of the system (right number), in MeV. }
\label{fig:ex-rho10}
\end{figure*}

\begin{table}[tb]
\caption{\label{table:constants} Values of adjustable model parameters. $\sqrt{\beta_{\Lambda K^*}}A_{j}$ represents the reduced helicity amplitude with $\beta_{\Lambda K^*}$ denoting the branching ratio for the resonance decay to $\Lambda K^*$,  and $A_j$ standing for helicity amplitude for the resonance with spin $j$ radiative decay to $\gamma p$. The asterisks ($\ast$) below resonance names denote the overall status of these resonances evaluated by PDG \cite{Tanabashi:2018}. The numbers in brackets below the resonance masses and widths represent the corresponding values estimated by PDG \cite{Tanabashi:2018}.}
\renewcommand{\arraystretch}{1.2}
\begin{tabular*}{\columnwidth}{@{\extracolsep\fill}lcc}
\hline \hline
$g^{(1)}_{\Sigma^\ast\Lambda\gamma}$  &  $-1.47 \pm 0.09$                  \\
$\Lambda_{K}$ [MeV]                		 	 &   $ 1009 \pm 2$                   	\\
\hline
$N^\ast$~Name                 			&  $N(2000){5/2}^+$    &  $N(2060){5/2}^-$ 	\\
                                                          &      $\ast\ast$          &    $\ast\!\ast\!\ast$    \\
$M_R$ [MeV]                  			&  $2010\pm 1$          &  $2043 \pm 12$          \\
								&                               &  $[2030 \sim 2200]$  \\
$\Gamma_R$ [MeV]             		       &   $400 \pm 95$          &   $79 \pm 3$         \\
								&				    &   $[300\sim 450]$      \\
$\Lambda_R$ [MeV]                		&   $1019\pm 6$          &   $1304\pm 8$         \\
$\sqrt{\beta_{\Lambda K^\ast}}A_{1/2}[10^{-3}{\rm GeV}^{-1/2}]$ & $-0.04 \pm 0.01$  & $0.55 \pm 0.03$    \\
$\sqrt{\beta_{\Lambda K^\ast}}A_{3/2}[10^{-3}{\rm GeV}^{-1/2}]$ & $0.59 \pm 0.02$  & $-1.40 \pm 0.05$      \\
$g^{(2)}_{R\Lambda K^\ast}/g^{(1)}_{R\Lambda K^\ast}$    & $-1.93 \pm 0.01$   &  $-1.96 \pm 0.24$  \\
$g^{(3)}_{R\Lambda K^\ast}/g^{(1)}_{R\Lambda K^\ast}$    & $-0.43 \pm 0.02$  &  $1.80 \pm 0.29$  \\
\hline \hline
\end{tabular*}
\end{table}

\begin{figure*}[tbp]
\includegraphics[width=0.8\textwidth]{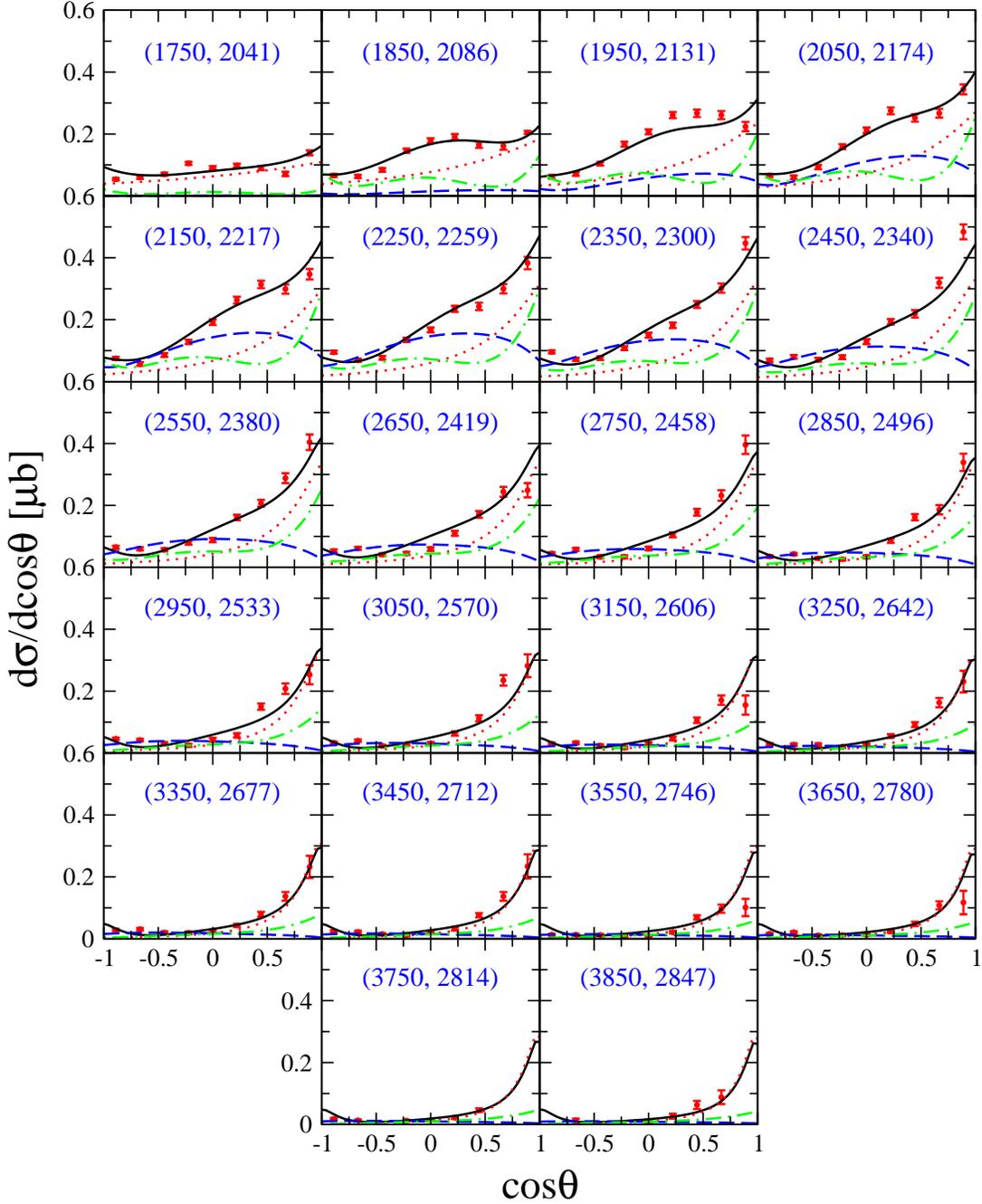}
\caption{Differential cross sections for $\gamma p \to K^{*+}\Lambda$ as a function of $\cos\theta$ in the center-of-mass frame (black solid line). The red dotted, blue dashed, and green dash-dotted lines represent the individual contributions from the $t$-channel $K$ exchange, the $s$-channel $N(2000)5/2^+$ exchange, and the $s$-channel $N(2060)5/2^-$ exchange, respectively. The data are taken from the CLAS Collaboration \cite{Tang:2013}. The numbers in parentheses, respectively, denote the photon laboratory incident energy (left number) and the total center-of-mass energy of the system (right number), in MeV.}
\label{fig:dsig}
\end{figure*}

\begin{figure*}[tbp]
\includegraphics[width=0.8\textwidth]{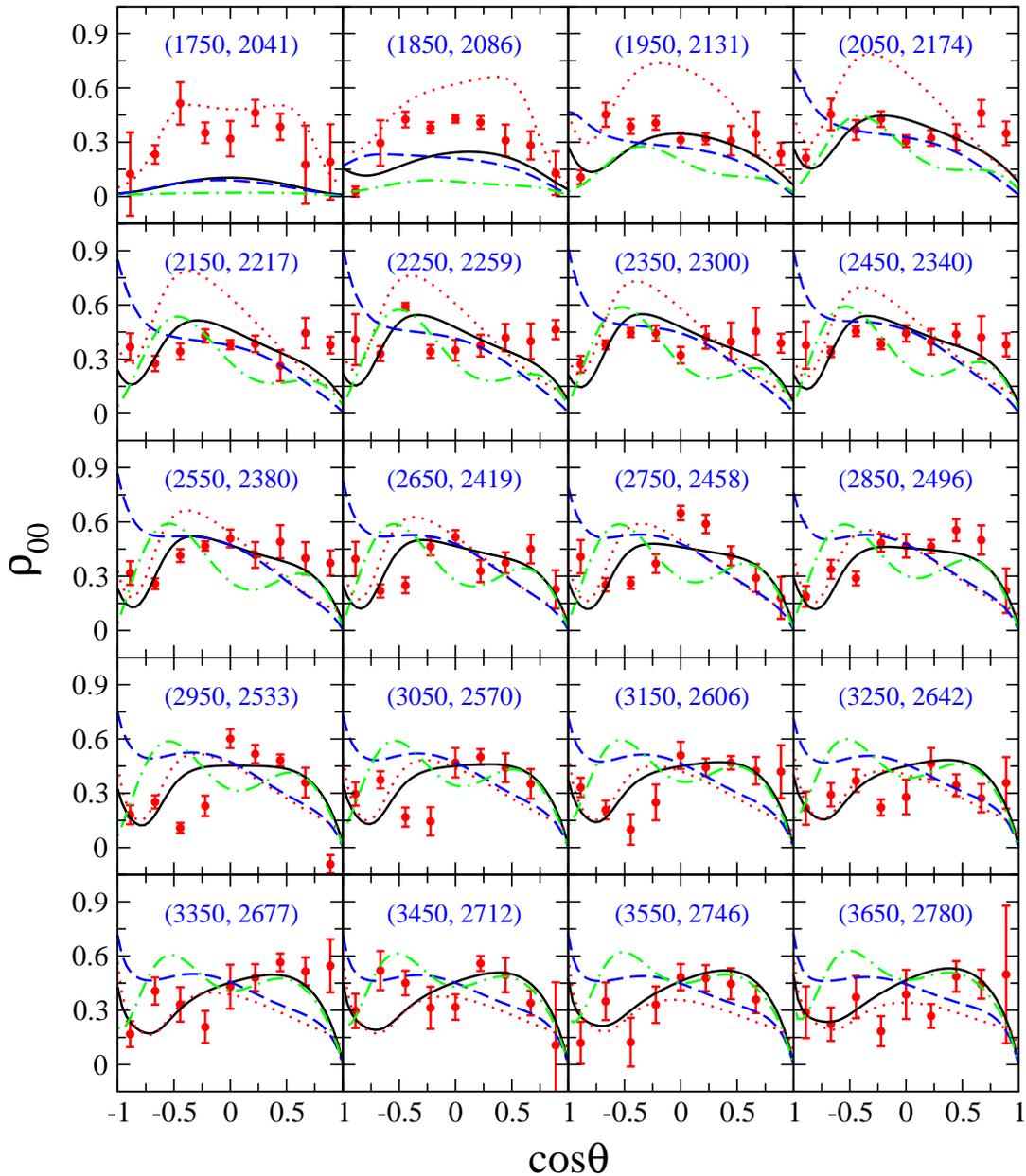}
\caption{Spin density matrix elements of $\rho_{00}$ for $\gamma p \to K^{*+}\Lambda$ as a function of $\cos\theta$ in the center-of-mass frame (black solid line). The red dotted, blue dashed, and green dash-dotted lines represent the results with the $t$-channel $K$ exchange, the $s$-channel $N(2000)5/2^+$ exchange, and the $s$-channel $N(2060)5/2^-$ exchange being switched off, respectively. The data are taken from the CLAS Collaboration \cite{Anisovich:2017rpe}. The numbers in parentheses, respectively, denote the photon laboratory incident energy (left number) and the total center-of-mass energy of the system (right number), in MeV.}
\label{fig:rho00}
\end{figure*}

\begin{figure*}[tbp]
\includegraphics[width=0.8\textwidth]{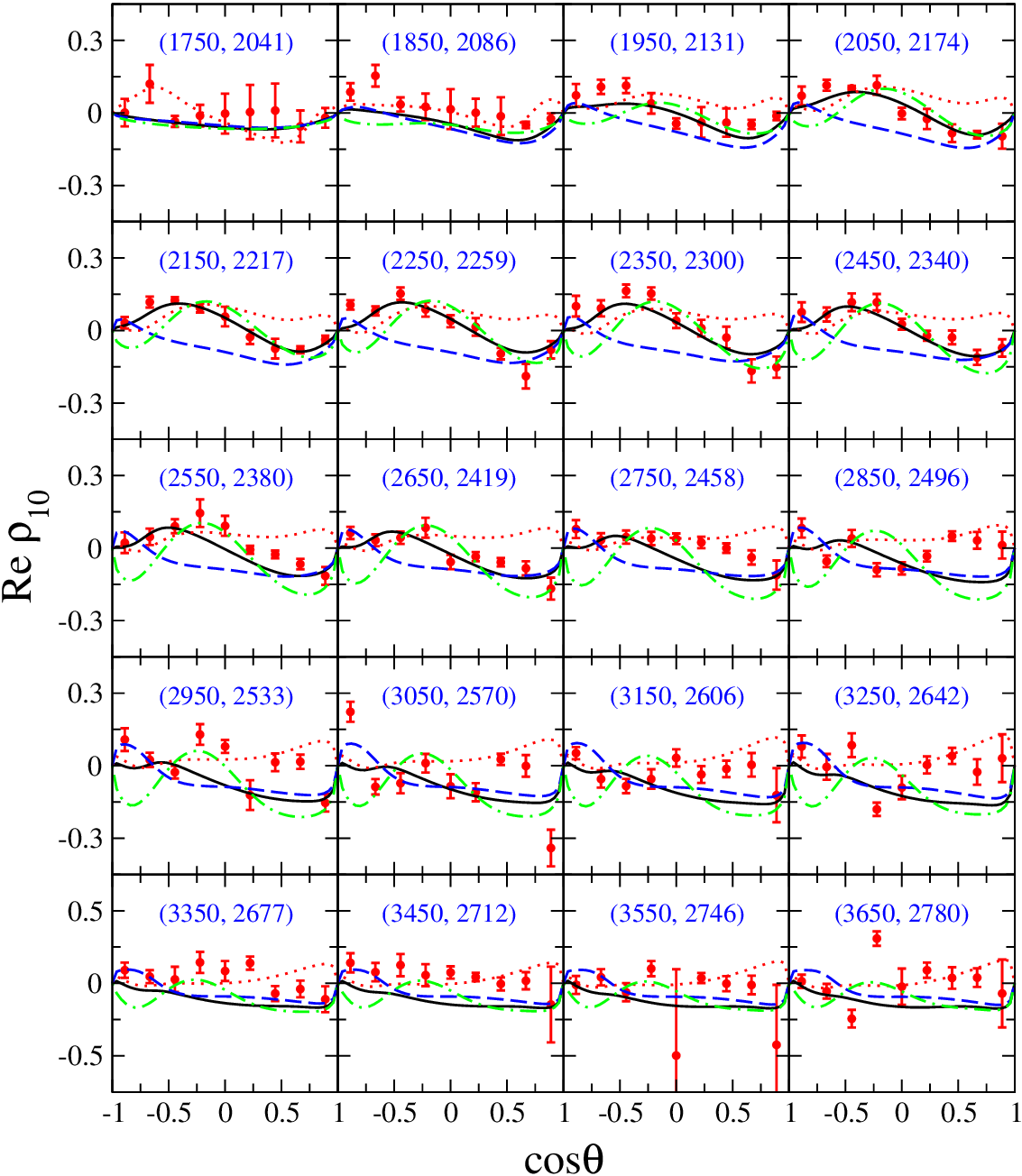}
\caption{Spin density matrix elements of ${\rm Re}\,\rho_{10}$ for $\gamma p \to K^{*+}\Lambda$ as a function of $\cos\theta$ in the center-of-mass frame (black solid line). Notations are the same as Fig.~\ref{fig:rho00}.}
\label{fig:rho10}
\end{figure*}

\begin{figure*}[tbp]
\includegraphics[width=0.8\textwidth]{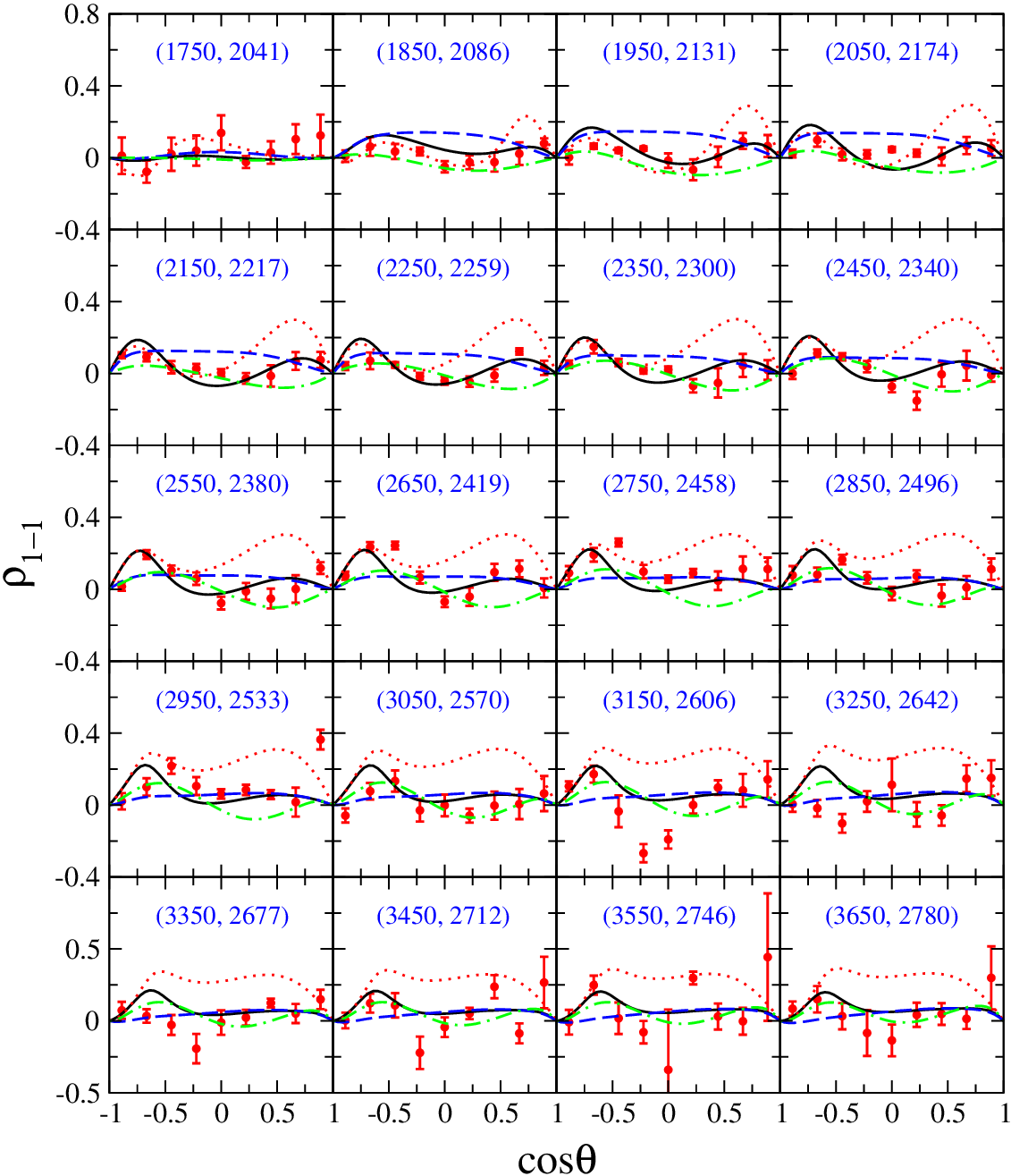}
\caption{Spin density matrix elements of $\rho_{1-1}$ for $\gamma p \to K^{*+}\Lambda$ as a function of $\cos\theta$ in the center-of-mass frame (black solid line). Notations are the same as Fig.~\ref{fig:rho00}.}
\label{fig:rho1m1}
\end{figure*}

Our previous work \cite{Wang:2017tpe} was devoted to employ an effective Lagrangian approach to analyze the high-precision differential cross-section data for $\gamma p \to K^{*+}\Lambda$ only, since at that time no data on spin observables were available. It was found in Ref.~\cite{Wang:2017tpe} that apart from the non-resonant contributions, one needs to introduce at least two nucleon resonances in the $s$ channel in constructing the reaction amplitudes in order to get a satisfactory description of the differential cross-section data. One of the needed resonances is $N(2060)5/2^-$, and the other one could be one of the $N(2000)5/2^+$, $N(2040)3/2^+$, $N(2100)1/2^+$, $N(2120)3/2^-$, and $N(2190)7/2^-$ resonances. With different choices of the other resonance besides $N(2060)5/2^-$, one got five fits with roughly similar fitting qualities, all visually in good agreement with the differential cross-section data. As the data on spin density matrix elements recently became available, it is natural to ask whether these new data can be automatically described by those five solutions resulted from the fits to the differential cross-section data or not. This has been carefully checked, and the answer is no. As an example, we show in Fig.~\ref{fig:ac} a comparison of the predictions of spin density matrix elements of $\rho_{1-1}$ from those five fits reported in Ref.~\cite{Wang:2017tpe} with the corresponding data at some selected energies. There, the five curves represent the predictions of $\rho_{1-1}$ from those five fits, and the scattered symbols are recent data from the CLAS Collaboration \cite{Anisovich:2017rpe}. It is clearly seen that there are big discrepancies  between the theoretical predictions and the experimental data, especially at the backward angles, and none of these five fits can satisfactorily describe the $\rho_{1-1}$ data. We have also tried to refit the model parameters by including the new data on spin density matrix elements in our data base, but it turns out that there is no way to get a simultaneous description of the data on both the differential cross sections and the spin density matrix elements.

Of course a good description of the data could in principle be achieved if one introduces a plentiful number of resonances into the reaction, since usually models with more resonances would have more adjustable parameters. However, this is not practical as the currently available data for $\gamma p\to K^{*+}\Lambda$ are far from enough to fully constrain a theoretical model. Actually, as has been discussed in Ref.~\cite{Pichowsky:1994gh}, in the case that the data have vanishing error bars, a complete determination of the 12 complex amplitudes for vector meson photoproduction reactions requires 23 independent observables (allowing for an overall arbitrary phase) at each energy and angle. The number of independent observables required for a complete determination of the amplitudes increases when the error bars are considered for real-world data. See Ref.~\cite{Nys:2015} for a discussion of the issue of completeness for pseudoscalar meson photoproduction. For the $\gamma p \to K^{*+}\Lambda$ reaction, it is obvious that the so far existing data on differential cross sections and spin density matrix elements are far from enough to uniquely determine the reaction amplitudes. If a lot of resonances are introduced in a model, the number of model parameters may be too many to be well constrained by the available data. In view of this, in the present work we take the same strategy as adopted in our previous work \cite{Wang:2017tpe} to introduce the nucleon resonances in constructing the $s$-channel reaction amplitude, i.e., we introduce the nucleon resonances as few as possible to describe the available data.

In Ref.~\cite{Wang:2017tpe}, the coupling constants $g_{R\Lambda {K^*}}^{(2)}$ and $g_{R\Lambda {K^*}}^{(3)}$ in the Lagrangians for the couplings of resonances with spin $J\geqslant 3/2$ to $\Lambda K^*$ are preset to zero in consideration of the fact that they can not be well determined with the differential cross-section data alone. Since the high precision data on spin density matrix elements for $K^*$ meson are now available, it is more reasonable to treat $g_{R\Lambda {K^*}}^{(2)}$ and $g_{R\Lambda {K^*}}^{(3)}$ as model parameters and let them being fixed by fitting all the available data.

As mentioned and discussed in Ref.~\cite{Wang:2017tpe}, the CLAS high-precision differential cross-section data \cite{Tang:2013} clearly show structures near the $K^*\Lambda$ threshold, indicating potential contributions from nucleon resonances in this energy region. In the most recent Particle Data Group (PDG) review \cite{Tanabashi:2018}, there are six nucleon resonances near the $K^*\Lambda$ threshold, namely, $N(2000)5/2^+$, $N(2040)3/2^+$, $N(2060)5/2^-$, $N(2100)1/2^+$, $N(2120)3/2^-$, and $N(2190)7/2^-$. It has been reported in Ref.~\cite{Wang:2017tpe} that if only one of these resonances is considered, even the differential cross-section data alone cannot be well described. When two resonances in the $s$ channel are introduced, one gets $15$ possible combinations, among which five are reported in Ref.~\cite{Wang:2017tpe} to be able to describe the differential cross-section data with similar fitting qualities. Here since for resonances with spin $J\geqslant 3/2$ we have two new terms in the resonance-$\Lambda$-$K^*$ vertices, i.e., the $g_{R\Lambda {K^*}}^{(2)}$ and $g_{R\Lambda {K^*}}^{(3)}$ terms, we will try all the 15 combinations of these six resonances to see if the data can be satisfactorily described. 

We use {\footnotesize MINUIT} to fit the model parameters, and have performed numerous tests by using various starting values of the parameters. In Table~\ref{table:chi2} the $\chi$ squared per data points for different types of observables for the best five fits are listed. There, in the second column, the numbers in the brackets denote the corresponding values for the data in the center-of-mass energy range $W<2217$ MeV. There are only $45$ data points for the differential cross sections in this energy region, which clearly exhibit apparent structures in the angular distributions indicating potential contributions from the nucleon resonances. Failing to describe these data may lead to low reliability of the extracted resonance information. We thus list the $\chi$ squared for the differential cross-section data in this energy region separately, which serves as a criterion to exclude the fits that fail to describe the differential cross-section data in this energy region but the chi squared for other observables in the whole energy region are comparable to the acceptable fits. 

From Table~\ref{table:chi2}, one sees that the obtained values of the $\chi$ squared for all the considered fits are relatively large, this is mainly due to the very small statistical uncertainties in the data. The solution with the $N(2040)3/2^+$ and $N(2120)3/2^-$ resonances has the smallest global $\chi^2/N$. However, it has $\chi^2_{d\sigma}/N_{d\sigma}=2.0$ for the differential cross-section data in the energy region $W\leq 2217$ MeV. Consequently, the shapes of the angular distributions near the $K^{*+}\Lambda$ threshold exhibited by the CLAS data cannot be described, as illustrated in Fig.~\ref{fig:compare}. There the dashed lines represent the results from the fit with the $N(2040)3/2^+$ and $N(2120)3/2^-$ resonances. One sees that the near-threshold structures in the angular distributions exhibited by the data are totally missed in this fit. Therefore, it is natural to exclude this solution as an acceptable fit. The solutions with the resonance $N(2040)3/2^+$ or $N(2120)3/2^-$ apart from $N(2060)5/2^-$ are excluded as acceptable fits by the same reason, since they have relative large $\chi^2_{d\sigma}/N_{d\sigma}$ ($1.9$ and $2.6$, respectively) for the differential cross-section data in the energy region $W\leq 2217$ MeV. The solution with the resonances $N(2000)5/2^+$ and $N(2060)5/2^-$ is taken as an acceptable fit. It can satisfactorily describe the overall data, which will be discussed later in detail. The solution with the $N(2190)7/2^-$ and $N(2060)5/2^-$ resonances has a global $\chi^2/N$ $15\%$ larger than that of our preferred fit, and moreover, it has a $\chi^2_{\rho_{10}}/N_{\rho_{10}}$ $36\%$ larger than that of our preferred fit. As an illustration, we show in Fig.~\ref{fig:ex-rho10} the predictions of spin density matrix elements of ${\rm Re}\,\rho_{10}$ from the solution with the $N(2190)7/2^-$ and $N(2060)5/2^-$ resonances (dashed lines) and our preferred fit with the $N(2000)5/2^+$ and $N(2060)5/2^-$ resonances (solid lines) compared with the corresponding data (scattered symbols) at some selected energies. One sees clearly that the results with the inclusion of the $N(2000)5/2^+$ and $N(2060)5/2^-$ resonances are in qualitative agreement with the data, while on the contrary, there are obviously big discrepancies between the theoretical results with the inclusion of the $N(2190)7/2^-$ and $N(2060)5/2^-$ resonances and the experimental data. Thus the solution with the $N(2190)7/2^-$ and $N(2060)5/2^-$ resonances is not considered as an acceptable fit. 

The fits with other combinations of two nucleon resonances result in larger values of global $\chi^2/N$ than those listed in Table~\ref{table:chi2}, and moreover, they either have $\chi^2_{d\sigma}/N_{d\sigma} \approx 2.0$ for the differential cross-section data in the energy region $W\leq 2217$ MeV, leading to unsuccessful descriptions of the shapes of the angular distributions near the $K^{*+}\Lambda$ threshold exhibited by the CLAS data, or have $\approx 35\%$ larger $\chi$ squared per data points for $\rho_{00}$, $\rho_{1-1}$ or ${\rm Re}\,\rho_{10}$, giving rise to much worse descriptions of the data on spin density matrix elements than our preferred fit, i.e., the fit with the $N(2000)5/2^+$ and $N(2060)5/2^-$ resonances. Therefore, these fits are excluded either to be acceptable fits.

Adding a third resonance to the model will bring in much more adjustable parameters. Consequently, too many solutions with similar fitting qualities can be obtained with large error bars for the fitting parameters, and no conclusive conclusion can be drawn about the resonance contents and parameters extracted from the available data for the considered reaction. We thus postpone the analysis with three or more resonances until more data become available in the future. 

We now concentrate on the discussions of the acceptable fit, which includes the $N(2000)5/2^+$ and $N(2060)5/2^-$ resonances in the construction of the reaction amplitudes, and results in a satisfactory description of all the available data. Note that this resonance set is the same one that results in the best description of the differential cross-section data in Ref.~\cite{Wang:2017tpe}. As mentioned above, the difference is that in the present work the $g_{R\Lambda {K^*}}^{(2)}$ and $g_{R\Lambda {K^*}}^{(3)}$ terms are allowed, while in Ref.~\cite{Wang:2017tpe} they are omitted. The values of all the adjustable model parameters in the present work are listed in Table~\ref{table:constants}, and the corresponding results for differential cross sections and spin density matrix elements of $\rho_{00}$, ${\rm Re}\,\rho_{10}$, and $\rho_{1-1}$ are shown in Figs.~\ref{fig:dsig}--\ref{fig:rho1m1}, respectively.

In Table~\ref{table:constants}, $M_R$, $\Gamma_R$, and $\Lambda_R$ denote the resonance mass, width and cutoff values, respectively. The asterisks below the resonance names represent the overall status of these resonances evaluated in the most recent review by PDG \cite{Tanabashi:2018}, and the numbers in brackets below the resonance mass and width are the corresponding values estimated by PDG.  $\sqrt{\beta_{\Lambda K^*}}A_{j}$ represents the reduced helicity amplitude for resonance, with $\beta_{\Lambda K^*}$ denoting the branching ratio for the resonance decay to $\Lambda K^*$ and $A_j$ standing for the helicity amplitude with spin $j$ for the resonance radiative decay to $\gamma p$. Note that as a common feature of a single channel analysis, the $s$-channel (resonance) amplitudes are sensitive only to the product of the hadronic and electromagnetic coupling constants. From Table~\ref{table:constants}, one sees that for the three-star resonance $N(2060)5/2^-$, the fitted mass $2043$ MeV is in the range of the PDG values $2030$--$2200$ MeV, and is also very close to the value reported in Ref.~\cite{Wang:2017tpe}, $2033$ MeV. The fitted width of this resonance, $79$ MeV, is much smaller than the PDG values, but is still very close to the value reported in Ref.~\cite{Wang:2017tpe}, $65$ MeV. One also observes that the reduced helicity amplitudes for $N(2060)5/2^-$ obtained from this work are also very close to the corresponding values reported in Ref.~\cite{Wang:2017tpe}. The reason that all the fitted mass, width and the reduced helicity amplitudes of $N(2060)5/2^-$ keep almost unchanged when the data on spin density matrix elements are taken into account is that the $N(2060)5/2^-$ resonance is responsible for the shape of the near-threshold angular distributions, which will be discussed later in detail. For the two-star $N(2000)5/2^+$ resonance, the fitted mass and width are $2010$ and $400$ MeV, respectively, both smaller than the values reported in Ref.~\cite{Wang:2017tpe}, $2115$ and $450$ MeV. The fitted value of the cutoff mass for the $t$-channel $K$ exchange is $\Lambda_K=1009$ MeV, which is very close to the value $1000$ MeV reported in Ref.~\cite{Wang:2017tpe}. The reason is that this parameter is mainly determined by the differential cross-section data at high energies where the $t$-channel $K$ exchange plays a dominant role.

Figure~\ref{fig:dsig} shows the results for the differential cross sections together with the contributions from the individual interaction diagrams. There, the black solid lines represent the contributions from the full amplitudes, and the red dotted, blue dashed, and green dash-dotted lines denote the contributions from the $t$-channel $K$ meson exchange, the $s$-channel $N(2000)5/2^+$ exchange, and the $s$-channel $N(2060)5/2^-$ exchange, respectively. The contributions from other terms are too small to be plotted. One sees that the overall theoretical differential cross sections are in qualitative agreement with the data. In the whole energy region considered, the dominant contribution comes from the $t$-channel $K$ exchange. In particular, the $K$ exchange is crucial to reproduce the observed forward-peaked angular distributions at higher energies. This also explains why in the present work the fitted cutoff mass for the $K$ exchange, $1009$ MeV, is very close to the value reported in Ref.~\cite{Wang:2017tpe}, $1000$ MeV. Near the threshold,  both the $s$-channel $N(2000)5/2^+$ and $N(2060)5/2^-$ exchanges have significant contributions, and the contribution from the $N(2060)5/2^-$ exchange is responsible for the shape of the angular distributions exhibited by the data. This feature has also been observed in Ref.~\cite{Wang:2017tpe}. The difference is that in the present work the contribution from the $N(2060)5/2^-$ resonance is stronger than that in Ref.~\cite{Wang:2017tpe} due to a lager cutoff value, and the shape of the differential cross sections resulted from the $N(2000)5/2^+$ resonance is different in these two works due to different fitted values of the mass, width, reduced helicity amplitudes, and cutoff mass for this resonance. The fact that the contribution of the $N(2060)5/2^-$ resonance is responsible for the shape of the near-threshold angular distributions explains why the fitted mass, width, and reduced helicity amplitudes for $N(2060)5/2^-$ in the present work are very close to those in Ref.~\cite{Wang:2017tpe}.

\begin{figure}[tbp]
\includegraphics[width=\columnwidth]{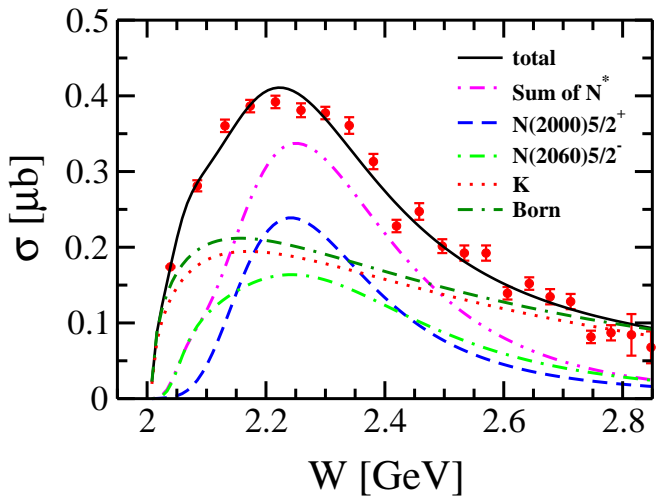}
\caption{Total cross sections with individual contributions for $\gamma p \to K^{*+}\Lambda$ as a function of the center-of-mass energy of the system. The data are taken from the CLAS Collaboration \cite{Tang:2013} but not included in the fit.}
\label{fig:total}
\end{figure}

In Figs.~\ref{fig:rho00}--\ref{fig:rho1m1}, we show the results for spin density matrix elements $\rho_{00}$, ${\rm Re}\,\rho_{10}$, and $\rho_{1-1}$. There, the black solid lines demonstrate the results from the full calculation, while the red dotted, blue dashed, and green dash-dotted lines denote the results calculated by switching off the contributions from the $t$-channel $K$ exchange, the $s$-channel $N(2000)5/2^+$ exchange, and the $s$-channel $N(2060)5/2^-$ exchange, respectively. As one can see, the overall results are in qualitative agreement with the recent CLAS data, even though some noticeable discrepancies are seen for all $\rho_{00}$, ${\rm Re}\,\rho_{10}$, and $\rho_{1-1}$. One also sees that the contributions from all the $t$-channel $K$ exchange and the $s$-channel $N(2000)5/2^+$ and $N(2060)5/2^-$ exchanges are rather important to the spin density matrix elements of $\rho_{00}$, ${\rm Re}\,\rho_{10}$, and $\rho_{1-1}$ in the whole energy regions considered.

Figure~\ref{fig:total} shows the predicted total cross sections (black solid line) together with the individual contributions from the $t$-channel $K$ exchange (red dotted line), the $s$-channel $N(2000)5/2^+$ exchange (blue dashed line), the $s$-channel $N(2060)5/2^-$ exchange (green dash-dotted line), the sum of both resonance exchanges (purple dash-double-dotted line), and the Bonn term (dark green double-dash-dotted line) which consists of the coherent sum of all the contributions other than the $s$-channel resonance exchanges. These quantities are obtained by integrating the corresponding results for differential cross sections. Note that the total cross section data are not included in our fit. One sees from Fig.~\ref{fig:total} that our predicted total cross sections are in fairly good agreement with the data over the entire energy region considered. The $t$-channel $K$ exchange is seen to play a significant role in the whole energy region stated, especially at high energies. The contributions from the Bonn term are very close to those from the $t$-channel $K$ exchange, indicating that the contributions from the non-resonant terms other than the $K$ meson exchange are rather small. The contributions from both the $N(2000)5/2^+$ and $N(2060)5/2^-$ exchanges are seen to be very significant, and the bump structure exhibited by the total cross section data is dominated by the coherent sum of the considered two resonances. Comparing Fig.~\ref{fig:total} with the corresponding total cross-section results from Ref.~\cite{Wang:2017tpe}, one sees that the contributions from the $t$-channel $K$ exchange in the present work are almost the same as those in Ref.~\cite{Wang:2017tpe}, with the reason being that the $K$ meson exchange is well constrained by the data on the angular distributions at high energies. The $s$-channel $N(2060)5/2^-$ exchange results in similar contributions at low energies in both the present work and the Ref.~\cite{Wang:2017tpe}, but at high energies this resonance has rather broader contributions in the present work due to a fitted larger cutoff mass. The contributions from the $s$-channel $N(2000)5/2^+$ exchange in the present work are quite different from those in Ref.~\cite{Wang:2017tpe}, as has been discussed in connection with the differential cross sections.

\section{Summary and conclusion}  \label{sec:summary}

In our previous work \cite{Wang:2017tpe}, we have analyzed the high-precision differential cross-section data for $\gamma p \to K^{*+}\Lambda$ reported by the CLAS Collaboration within an effective Lagrangian approach. There, we have considered the $t$-channel $K$, $K^*$, and $\kappa$ exchanges, the $u$-channel $\Lambda$, $\Sigma$, and $\Sigma^*$ exchanges, the interaction current, the $s$-channel $N$ exchange and nucleon resonances exchanges in constructing the reaction amplitudes. It was found that in order to describe the differential cross-section data, one needs to introduce in the $s$ channel at least two nucleon resonances. One of the needed resonances is the $N(2060)5/2^-$, while the other one can not be uniquely determined. It could be any one of the following five resonances: $N(2000)5/2^+$, $N(2040)3/2^+$, $N(2100)1/2^+$, $N(2120)3/2^-$, and $N(2190)7/2^-$. As a consequence, five different fits were obtained with roughly similar fitting qualities, all visually in good agreement with the cross-section data.

Recently the data on spin density matrix elements for $\gamma p \to K^{*+}\Lambda$ have been reported by the CLAS Collaboration \cite{Anisovich:2017rpe}. As far as we know, these new data have so far never been analyzed in any theoretical works. In this paper, we for the first time include these data into our analysis within an effective Lagrangian approach based on our previous work \cite{Wang:2017tpe}. The purpose is to impose further constraints on extracting the resonance contents and the associated resonance parameters, and to gain a better understanding of the reaction mechanism for this photoproduction reaction.

Our results show that a simultaneous and satisfactory description of the data on both the spin density matrix elements and the differential cross sections can be achieved by introducing in the $s$ channel two nucleon resonances, namely the $N(2060)5/2^-$ and $N(2000)5/2^+$ resonances. Further analysis shows that this reaction is dominated by the $t$-channel $K$ exchange and the $s$-channel $N(2060)5/2^-$ and $N(2000)5/2^+$ exchanges. The $K$ meson exchange is crucial to explain the forward-peaked feature of the differential cross sections in the high-energy region. The $N(2060)5/2^-$ resonance exchange is responsible for the shape of the near-threshold angular distributions exhibited by the data. The mass, width, and reduced helicity amplitudes for the three-star resonance $N(2060)5/2^-$ fitted in the present work are very close to those reported in our previous work \cite{Wang:2017tpe}. The parameters of the two-star resonance $N(2000)5/2^+$ obtained in the present work are found to be quite different from those reported in Ref.~\cite{Wang:2017tpe}. More data on spin observables would be helpful to further constrain the resonance contents and the associated resonance parameters extracted from this reaction.

\begin{acknowledgments}
This work is partially supported by the National Natural Science Foundation of China under Grants No.~11475181 and No.~11635009, the Youth Innovation Promotion Association of CAS under Grant No.~2015358, and the Key Research Program of Frontier Sciences of CAS under Grant No.~Y7292610K1.
\end{acknowledgments}

\end{document}